\documentstyle[12pt,epsf]{article}
\newcommand{\br}{{\bf r}}
\newcommand{\half} {\frac{1}{2}}

\newcommand{\lk} {l_k}

\newcommand{\bsigma}{\mbox{\boldmath $\sigma$}}
\newcommand{\btau}{\mbox{\boldmath $\tau$}}
\newcommand{\threej}[6]{ \left( \begin{array}{ccc}
                               #1 & #2 & #3 \\
                               #4 & #5 & #6 
                             \end{array}
                        \right) } 
\newcommand{\sixj}[6]{ \left\{ \begin{array}{ccc}
                               #1 & #2 & #3 \\
                                #4 & #5 & #6 
                               \end{array}
                        \right\} } 
\newcommand{\ninej}[9]{ \left\{ \begin{array}{ccc}
                                 #1 & #2 & #3 \\
                                 #4 & #5 & #6 \\
                                 #7 & #8 & #9 
                                \end{array}
                         \right\} } 
\begin{document}
\centerline{\large 
{\bf Spin-orbit interaction in Hartree--Fock calculations}}
\vspace{1.5cm}
\centerline{\bf Ana R. Bautista}
\centerline{Departamento de F\'{\i}sica Moderna, 
Universidad de Granada,}
\centerline{E-18071 Granada, Spain}
\vspace{.5cm}
\centerline{\bf Giampaolo Co'}
\centerline{Dipartimento di Fisica, Universit\`a di Lecce}
\centerline{and I.N.F.N. sezione di Lecce, I-73100 Lecce, Italy}
\vspace{.5cm}
\centerline{\bf Antonio M. Lallena}
\centerline{Departamento de F\'{\i}sica Moderna, 
Universidad de Granada,}
\centerline{E-18071 Granada, Spain}
\vspace{1.5cm}
\begin{abstract}
The contribution of the spin-orbit interaction in Hartree--Fock
calculation for closed shell nuclei is studied.  We obtain explicit
expressions for the finite range spin-orbit force. New terms with
respect to the traditional spin-orbit expressions are found.  The
importance of the finite--range is analyzed.
Results obtained with spin-orbit 
terms taken from realistic interactions 
are presented. 
The effect of the spin--orbit isospin dependent terms is
evaluated.
\end{abstract}
\vskip 0.5 cm
PACS {21.60.Jz}
\newpage
\section{Introduction}
The success of the shell model in predicting the nuclear magic numbers
is related to the presence of a strong spin-orbit term in the nuclear
average potential.  Few years after the formulation of the nuclear
shell--model \cite{fee49}, evidences of spin-orbit terms in the
nuclear interaction were identified \cite{fer54} by analyzing
polarized proton scattering data off complex nuclei \cite{hen52}.

The strong spin--orbit term of the nuclear average potential should be
generated by an analogous term present in the nucleon--nucleon
interaction. However, the connection between realistic
nucleon--nucleon interactions, i.e. those built to reproduce the two
nucleon scattering data and the deuteron properties, and the effective
interactions, those used in nuclear structure effective theories, is
still unclear.

In the present article we present a study on the relationship between
the spin--orbit terms used in realistic nucleon--nucleon interaction
and the analogous terms used in Har\-tree--Fock (HF) calculations.  Our work
has been done within the non relativistic framework where the
spin--orbit terms can be easily isolated.  In effect, in our
HF calculations, we have used the explicit spin-orbit terms
of the Argonne--Urbana nucleon-nucleon realistic potentials
\cite{lag81,wir84,wir95} describing, within the non relativistic
framework, nucleon--nucleon elastic scattering data up to energies of
about 300 MeV.

The spin--orbit interactions commonly used in HF
calculations are of zero--range type and, in general, they 
are parametrized following the expressions
proposed by Skyrme \cite{bel56}.  Also the Gogny interaction
\cite{gog80}, which has a finite range for all the other channels, uses a
Skyrme--like expression for the spin--orbit term.

We explicitly develop the expressions for a finite range spin--orbit
interaction to be used in HF calculations. As expected, in
addition to the direct term these expressions produce a contribution
also in the Fock--Dirac exchange term of the HF
equations. This term is not present when zero--range interactions are
used. Also in the direct term there are some new contributions with
respect to the expression obtained with the Skyrme interaction.

In the next paragraphs we present the detailed expressions of the
HF equations when finite range spin--orbit terms are
considered, then we discuss the importance of the finite--range and
the effect of using spin--orbit terms taken from realistic
nucleon--nucleon interactions. Finally we draw our conclusions.

\section{The formalism}
In ref.~\cite{co98} we made an explicit presentation of the
HF formalism with finite--range interactions. In the
present article we extend the formalism in order to treat also the
spin--orbit terms. For this reason we recall here only those parts
of the formalism involving the spin-orbit terms.

The effective interaction used in our calculations has the form:
\begin{equation}
\label{force1}
V(\br_1,\br_2) \, = \, \sum_{p=1}^8 \, V_p(\br_1,\br_2) \, O^p(1,2)
\end{equation}
where to the first 6 components used in  ref.~\cite{co98} we have added
the spin--orbit terms defined as $O^7(1,2)={\bf L} \cdot {\bf S}$ and
 $O^8(1,2)={\bf L} \cdot {\bf S} \,\, \btau_1 \cdot \btau_2$
with 
${\bf L}= ({\bf r}_1-{\bf r}_2)\times({\bf p}_1-{\bf p}_2)$ 
and ${\bf S}=\half (\bsigma_1 + \bsigma_2)$.

To take advantage of the spherical symmetry of the problem, we
describe the single particle wave functions by separating the angular
and the radial parts 
$\phi_k(\br)\equiv\tilde{k}(\Omega) u_k(r)/r$
where the subindex $k$ indicates all the quantum numbers necessary to
identify the state and $\Omega$ the two angular coordinates $\theta$
and $\phi$.

The specific expression of the single particle wave functions allows
us to reduce the HF equations into a set of differential equations of
the type:
\begin{equation}
\label{hf6}
-\frac{\hbar^2}{2m_k}\left(\frac{{\rm d}^2}{{\rm d}r^2}-
\frac{\lk(\lk+1)}{r^2}\right)
u_k(r) + U_k(r)u_k(r) - W_k(r) = \epsilon_k u_k(r),
\end{equation}
\noindent
where we have defined:
\begin{equation}
\label{ukr}
U_k(r)\, =\, \sum_i \int {\rm d}r' \, u^*_i(r')\, 
\int {\rm d} \Omega  \int {\rm d} \Omega ' \,\,
\tilde{k}^* (\Omega) \tilde{i}^*  (\Omega ')
 V(|{\bf r}-{\bf r}'|)
\tilde{k} (\Omega) \tilde{i}  (\Omega') \,\,
 u_i(r'),
\end{equation}
and
\begin{equation}
\label{wkr}
W_k(r) \, = \, 
\sum_i \int {\rm d}r' \, u^*_i(r')\, 
\int {\rm d} \Omega  \int {\rm d} \Omega ' \,\,
\tilde{k}^* (\Omega) \tilde{i}^*  (\Omega ')
V(|{\bf r}-{\bf r}'|) 
\tilde{i} (\Omega) \tilde{k}  (\Omega') \,\,
u_i(r)\, u_k(r').
\end{equation}

While the interaction depends from the relative distance between two
nucleons, the HF equations (\ref{hf6}) depend upon 
the distance of the particles from the origin of the reference system. 
The implementation
of finite range interactions in the HF equations requires the
separation of the coordinate variables in the interaction.
For the central and tensor channels this separation is done by
considering the interaction in coordinate space as Fourier transform
of the interaction expressed in momentum space (see ref.~\cite{co98}
for details).
For the spin--orbit channels ($p=7,8$) we use a different strategy 
consisting in expanding in multipoles the interaction:
\begin{equation}
V_p(r_{12})=4\pi\, \sum_{LM}\frac{1}{\widehat{L}^2}
{\cal V}^p_L(r_1,r_2)\,
Y^*_{LM}(\widehat{r}_1)Y_{LM}(\widehat{r}_2) .
\end{equation}
\noindent
with $\widehat{L} \equiv \sqrt{2L+1}$.
>From the previous equation, making use of the orthogonality of the
spherical harmonics, we obtain a close expression for the coefficients
of the expansion:
\begin{equation}
\label{vlp}
{\cal V}^p_L(r_1,r_2)\,=\,\frac{\widehat{L}^2}{2}\,\int^1_{-1}\,
{\rm d}\cos \theta_{12}\,V_p(r_{12})\,P_L(\cos\theta_{12})\, ,
\end{equation}
In
the previous equation we have indicated with $P_L$ the Legendre
polynomials and with $\theta_{12}$ the angle between ${\bf r}_1$ and
${\bf r}_2$.

The details of the calculations of the spin--orbit matrix elements are
given in the Appendix. We obtain for the direct term in the HF
equations the following result:
\begin{equation}
\label{dls}
\left[U_k(r)\right]_{p=7,8}\,=\, 
2 \pi \, I^p_k\,\int {\rm d}r' r'^2
\left\{\left[j_k\left(j_k+1\right)-l_k\left(l_k+1\right)-
\frac{3}{4}\right]\,
{\cal U}^p_C(r,r')+{\cal U}^p_{LS}(r,r')\right\},
\end{equation}
where
\begin{equation}
\label{ipk}
I^p_k \, = \, \left\{ \begin{array}{ll}
                1,   & \mbox{$p$=7} \\
                2 \, t_k, & \mbox{$p$=8}\,\,\,\,\,\,, 
\end{array}
\right.
\end{equation}
and the potentials ${\cal U}^p_C$ and ${\cal U}^p_{LS}$ are given by:
\begin{equation}
\label{upc2}
{\cal U}^p_C(r,r')\,=\,\left[{\cal
V}^p_0(r,r')-\frac{1}{3}\frac{r'}{r}{\cal
V}^p_1(r,r')\right]\,\Omega^p_C(r'),\,\,\,\,\,\,\,\, p=7,8
\end{equation}
and
\begin{equation}
\label{upls}
{\cal U}^p_{LS}(r,r')\,=\,\left[{\cal
V}^p_0(r,r')-\frac{1}{3}\frac{r}{r'}
{\cal V}^p_1(r,r')\right]\,\Omega^p_{LS}(r'),\,\,\,\,\,\,\,\,p=7,8.
\end{equation}
The function $\Omega_C^p(r)$ used in the previous equations has been 
defined as:
\begin{equation}
\label{uuu}
\Omega^p_C(r) \, = \, \left\{ \begin{array}{ll}
        \rho(r) \, , & \mbox{$p$=7} \\ \\
        \rho^{\pi}(r)-\rho^{\nu}(r) \, , & \mbox{$p$=8}\, ,
\end{array}
\right.
\end{equation}
where $\rho^{\pi}(r)$ and $\rho^{\nu}(r)$ are the proton and neutron
densities such as  $\rho^{\pi}(r) + \rho^{\nu}(r)=\rho(r)$.
The other function used in eq. (\ref{upls}) has been
defined as:
\begin{equation}
\Omega^p_{LS}(r) \, = \, \left\{ \begin{array}{ll}
        \rho_{LS}(r) \, , & \mbox{$p$=7} \\ \\
        \rho^{\pi}_{LS}(r)-\rho^{\nu}_{LS}(r) \, , & 
\mbox{$p$=8}\, ,
\end{array}
\right.
\end{equation}
where $\rho_{LS}$ is the nucleon spin-density,
\begin{equation}
\label{lsden}
\rho_{LS}(r)=\frac{1}{4\pi}\sum_i
\left[j_i(j_i+1)-l_i(l_i+1)-\frac{3}{4}\right]\,
\widehat{j}^2_i\,\left(\frac{u_i(r)}{r}\right)^2
\, ,
\end{equation}
\noindent
and $\rho^{\pi}_{LS}(r)$ and $\rho^{\nu}_{LS}(r)$ 
are the analogous functions for protons and neutrons respectively.

For the  exchange terms of eq. (\ref{hf6}) we obtain:
\begin{equation}
\label{els}
\left[W_k(r)\right]_{p=7,8}\,=\, \sum_{iL} \, I^p_{ki}\,
\sum_{\alpha=1,5}\,\varepsilon^{(\alpha)}_{kiL}\,
\int\,{\rm d}r'\,{\cal W}^{p(\alpha)}_{kiL}(r,r')  \, ,
\end{equation}
where 
\begin{equation}
\label{ipki}
I^p_{ki}\, = \, \left\{ \begin{array}{ll}
\delta_{t_k,t_i}\,,   & \mbox{$p=7$} \\ \\
2\delta_{t_k,-t_i}+ \delta_{t_k,t_i}\,, & \mbox{$p=8$}  \,\,\, .
\end{array}
\right.
\end{equation} 
The new five functions  $\varepsilon$ have been defined as:
\begin{equation}
\varepsilon^{(\alpha)}_{kiL}\,=
\,\sqrt{3}\,(-1)^{j_i+l_i+\half}\,\widehat{l}_k\,
\widehat{l}_i\,\widehat{j}_i^2\,\sum_K(-1)^K\,\widehat{K}^2\,
\threej{1}{K}{L}{1}{-1}{0}\,\zeta^{(\alpha)}_{ki}(L,K),
\,\,\,\,\,\, \alpha=1,\ldots,5 \, ,
\end{equation}
with $\zeta^{(\alpha)}_{ki}(L,K)$ given by:
\begin{eqnarray*}
\zeta^{(1)}_{ki}(L,K)&=&\xi(l_k+l_i+L)\,{\cal T}_{ki}(L,K)
\nonumber\\ 
&\mbox{}&\,\,\,\,\,\,\,\,\,\,\,
\left[\sqrt{l_i(l_i+1)}\threej{l_i}{l_k}{K}{-1}{0}{1}
-\sqrt{l_k(l_k+1)}\threej{l_i}{l_k}{K}{0}{-1}{1}\right]
\\ \nonumber \\ \nonumber \\
\zeta^{(\alpha)}_{ki}(L,K)&=&{\cal G}_{ki}(L,K)\,
\left\{\begin{array}{ll} 
        (-2)\sqrt{l_i(l_i+1)}\threej{l_i}{l_k}{K}{-1}{0}{1}
\threej{1}{K}{L}{1}{-1}{0}, & \alpha=2 \\ \\
        2\sqrt{l_k(l_k+1)}
\threej{l_i}{l_k}{K}{0}{-1}{1}\threej{1}{K}{L}{1}{-1}{0}, & 
\alpha=3 \\ \\
        (-\sqrt{2})\threej{l_i}{l_k}{K}{0}{0}{0}
\threej{1}{K}{L}{0}{0}{0}, & \alpha=4,5 \, .
\end{array}
\right.
\nonumber
\end{eqnarray*}
\noindent
In the previous expressions we have used the following definitions:
\begin{eqnarray*}
{\cal T}_{ki}(L,K)&=&\ninej {l_k}{\frac{1}{2}}{j_k}
{l_i}{\frac{1}{2}}{j_i}{K}{1}{L} 
\threej {j_k}{j_i}{L}{\frac{1}{2}}{-\frac{1}{2}}{0} 
\nonumber 
\\ 
&\mbox{}&\,\,\,\,\,\,\,\,\,\,\,-\widehat{l}_k \widehat{l}_i
\ninej {l_k}{\frac{1}{2}}{j_k}{l_i}{\frac{1}{2}}{j_i}{L}{1}{K}
\sixj {l_k}{l_i}{K}{j_i}{j_k}{\frac{1}{2}}  
\threej {l_k}{l_i}{L}{0}{0}{0},\\
{\cal G}_{ki}(L,K)&=& \xi(l_k+l_i+L+1)\,
\sum_{L'}\,\xi(L+L'+1)\,\widehat{L'}^2\,
\threej{1}{K}{L'}{1}{-1}{0}\,{\cal T}_{ki}(L',K).
\nonumber
\end{eqnarray*}
In the eqs. (\ref{els}) we have used five new potentials:
\begin{eqnarray*}
{\cal W}^{p(\alpha)}_{kiL}(r,r')&=&u^*_i(r')\, {\cal
V}^p_L(r,r')\,u_k(r')u_i(r)\,\left\{\begin{array}{ll}
        1, & \alpha=1, \\\\
        \frac{\mbox{$r'$}}{\mbox{$r$}}, & \alpha=2, \\\\
        \frac{\mbox{$r$}}{\mbox{$r'$}}, & \alpha=3, 
\end{array}
\right.
\nonumber
\\ \\
{\cal W}^{p(\alpha)}_{kiL}(r,r')&=&\left\{\begin{array}{ll}
                r'\,u^*_i(r')u_k(r')\,{\cal V}^p_L(r,r')\,
\frac{\mbox{{\rm d}}}
{\mbox{{\rm d}}\mbox{$r$}}u_i(r), & \alpha=4, \nonumber\\\\
                r\,u_k(r')
\frac{\mbox{{\rm d}}}{\mbox{{\rm d}}\mbox{$r'$}}[u^*_i(r')\,{\cal
V}^p_L(r,r')]\,u_i(r), & \alpha=5. 
\end{array}
\right.
\end{eqnarray*}
Like in ref.~\cite{co98} 
the numerical solution of eq. (\ref{hf6}) has been obtained iteratively
using the plane wave expansion method of refs.~\cite{gua82}.
The center of mass motion has been considered in its simplest
approximation, consisting in inserting the nucleon reduced mass in the
hamiltonian. The single particle wave functions used to start the
iterative procedure have been generated by a Saxon--Woods potential
without spin--orbit and Coulomb terms. Therefore the starting wave
functions for spin--orbit partners are the same.

\section{Results}

In the same spirit of the work of  ref.~\cite{co98}
we are more interested in investigating the validity of 
the commonly used approximations rather than proposing
a new effective interaction to be used in HF calculations.
This study has been conducted by adding different kinds of spin--orbit
terms to a basic interaction composed by the four central terms of the
force. These terms are described as a sum of two gaussians:
\begin{equation}
V_p(r)=\sum_{i=1}^2 A_{pi} \, exp(-b_i r^2) \, , 
\end{equation}
with $p=1,2,3,4$. The parameters of this part of the interaction,
which we call $B1a$, are compared in tab. \ref{tb1b1a} with the
parameterization $B1$ of Brink and Boeker \cite{bri67}.  
The small differences are due to the fact that we have considered
the Coulomb interaction and therefore we had to readjust
the parameters of the force in order to reproduce the binding energy
of $^4$He.
We added to the $B1a$ interaction a finite range spin--orbit
term of gaussian form:
\begin{equation}
\label{sofin}
V_7(r_{12})=A_7 \, exp(-b_7 r^2_{12})  \, .
\end{equation}
The finite range effects have been investigated by comparing the
results obtained with the above interaction with those produced by
adding to the  $B1a$ force a zero range spin-orbit term of the
form:
\begin{equation}
\label{sozer}
V_7(r_{12})=A_7 \,\, \delta^3({\bf r_1} -{\bf r_2}) \, .
\end{equation}
The straightforward insertion of this expression in our formalism
gives a contribution exactly equal to zero. The reason of this result
can be traced back to the fact that we have developed our expressions 
using ${\bf L}=(\br_1- \br_2) \times ({\bf p}_1-{\bf p}_2)$.  To
get results different from zero for a  zero-range spin--orbit
interaction we set to zero the quantity ${\cal V}^p_1(r,r')$ in
eqs. (\ref{upc2}) and (\ref{upls}), and after
inserting eq. (\ref{sozer}) we obtained:
\begin{equation}
\label{upzero}
{\cal U}^7_{C,LS}(r,r')\,=\,
\frac{A_7}{2r} \,\,
\Omega^7_{C,LS}(r')\,\delta(r-r') \,.
\end{equation}
The calculations done with this approach are labelled as $z$.

In addition to these  effective interactions we have also used
spin--orbit terms taken from microscopic forces: 
the Urbana V14 \cite{lag81}, Argonne V14 \cite{wir84}, and
Argonne V18 \cite{wir95} potentials.  
Our study has been restricted
to the investigation of the doubly magic nuclei $^{12}$C, $^{16}$O,
$^{40}$Ca, $^{48}$Ca and $^{208}$Pb.

The finite range interaction (\ref{sofin}) has been used to study
the role played by the various terms of the spin--orbit potential.
In a first set of calculations, labelled as $c$, only the
${\cal U}_C$ term of eq. (\ref{dls}) has been used. This is the only
spin--orbit term present in shell--model calculations. 
In another set of calculations, denoted as $d$, we considered the full
direct term, and, finally, 
the results identified with $so$ have been obtained with all
the spin--orbit terms. 
These calculations have been done by changing every time the
parameters of the force  (\ref{sofin})
to reproduce  the 6.3 MeV splitting between the protons 1p levels in
$^{16}$O. 
The parameter $b_7$ was fixed to the arbitrary value of 1.2 fm$^{-2}$
and the fit of the splitting was obtained by changing $A_7$. 
The values of $A_7$ obtained in this way are, -108.75, -107.50, 
-97.86 MeV for
the $c$, $d$, and $so$ calculations respectively. The three
interactions do not differ very much as it is shown in the panel I of
the figure. This result indicates that the largest contribution from the
spin--orbit force is coming from the  ${\cal U}_C$ factor of the direct
term.
Since the other terms are small we have explored the possibility of
avoiding their explicit calculation by simulating their effects with a
readjustment of the force parameters. This is the reason why each
type of calculation has been done with a different parametrization of
the force, each of them reproducing the same empirical quantity.

In tab. \ref{be} we compare the binding energies obtained with our
calculations with the experimental ones \cite{aud93}. In
spite of the fact that we handle with a non--linear problem, the
spin--orbit terms acts on the binding energies as expected.  The main
contribution to the binding energy is obtained by the sum of the
single particle energies. In nuclei where all the spin--orbit partners
are occupied the spin--orbit term lowers the energy of the $l+1/2$
level and increases that of the $l-1/2$ level, in such a way that the
contribution to the nuclear binding energy is almost zero. 
In table \ref{be} this is observed by looking at the values of
the energies of $^{16}$O and $^{40}$Ca
which are practically the same, independently from the spin-orbit force
used.  Clearly those nuclei where not all the spin--orbit partners
are occupied are sensitive to the spin--orbit force, since the single
particle energy of the last occupied level is lowered. The effect is
seen in $^{12}$C, $^{48}$Ca and $^{208}$Pb where the binding energy
increases, in absolute value, the stronger the spin-orbit force is.

The quantity most sensitive to the spin--orbit
interaction is the energy splitting between spin--orbit partners
levels.  The energy splittings calculated for the various
nuclei under investigation with the interactions proposed
are compared in tabs. \ref{tz1} and \ref{tz2} with the Skyrme III
\cite{bei75} 
results and with the empirical values \cite{cam72}. The
experimental spectrum is more compressed than the theoretical one.  
This fact is well known \cite{rin80}, and it is
related to the intrinsic limitations of the HF theory in the
description of an interacting many--body system.

As expected, the
splittings increase with increasing value of $l$.  The splittings
obtained with the zero range interaction $z$ become larger than those
obtained with finite range interaction as the mass number of the
nucleus increases.  
The results obtained with zero--range Skyrme interaction 
do not present this effect.
In the Skyrme interaction there are velocity dependent terms 
generating spin--orbit like contributions which add to those 
produced by the genuine spin--orbit term.  
These velocity dependent terms simulate the effects of the finite range.  
We observe that the value of the splittings obtained
with the Skyrme III interaction are comparable with those
obtained with our finite range interactions.

>From the comparison of the results of the $c$ and $d$ columns of
tabs. \ref{tz1} and \ref{tz2} we infer information on the role of the
terms ${\cal U}^p_{LS}$ in eq. (\ref{dls}). 
The inclusion of ${\cal U}^p_{LS}$ increases the splitting
for all the nuclei considered but the magnitude of this increase is
rather different for the various nuclei. We should not consider in our
analysis the nucleus $^{16}$O since it has been used to fit the
interaction. 
We notice that the addition of ${\cal U}^p_{LS}$ produces quite small
differences in 
the splitting of $^{40}$Ca and $^{208}$Pb nuclei while they are
remarkable in $^{12}$C and $^{48}$Ca.

These results can be understood by considering that ${\cal U}^p_{LS}$
is related to the nuclear spin density, eq. (\ref{lsden}). If we
assume that the radial wave functions $u(r)$ are the same for
spin--orbit partners levels, the contribution of these two levels to
the spin density is exactly zero. In real calculations these wave
functions are slightly different, but the contribution to the spin
density remains small.  This explains the small increase of the
splitting in $^{40}$Ca and the relatively large modifications produced
in $^{12}$C and $^{48}$Ca. One should remark that only the unoccupied
levels contribute to the spin density. For this reason the effect of
${\cal U}^p_{LS}$ is relatively large with respect to that of ${\cal
U}^p_{C}$ in $^{12}$C and $^{48}$Ca where the number of single
particle levels is relatively small. In a heavy nucleus like $^{208}$Pb
there are many levels contributing in ${\cal U}^p_{C}$ and the effects
of ${\cal U}^p_{LS}$ produced by a single level is relatively small.

The contribution of the exchange term can be seen by comparing the
results of the $d$ and $so$ columns. 
The variations with respect to the calculations done with
only the direct terms can be as big as 10-15\%, but not all of them
have the same sign. It seems that for all the $p$ states the
splitting is reduced when the exchange term is considered, but it is
increased in the $f$, $g$ and $h$ states. The situation for the $d$
states is even more complicated, since the splitting is reduced for
the $1d$ states in $^{208}$Pb but it has increased for all the other
$d$ states. The contribution of the exchange term cannot be
taken into account in calculations with the direct term only by
modifying the force parameters. 

The values of the splittings produced by the Urbana (U), and
Argonne V$_{18}$ interactions are comparable with those of our
interactions, while the Argonne V$_{14}$ (A14) generates smaller values.
The radial dependence of the spin--orbit terms of these interactions
are shown in the panel II of the figure.
It is remarkable that the results of U and A18 are similar in spite of
the large difference in the depth. The depth value of A14 is
intermediate between those of the  previous two forces, but its
splittings are smaller. Th U and A18 forces have similar range, while
that of A14 is smaller.
These facts indicate that our calculations are more sensitive to
the range of the interaction than to its minimum value. In effect we
recall that, in our calculations, a zero--range interaction does
not produce any splitting.

The calculations done with the microscopic interactions
include both spin--orbit and spin--orbit isospin terms.   
In order
to study the importance of the isospin part of the spin--orbit
interaction we have repeated each calculation leaving out this terms.
The differences of the results obtained with the full interaction and
those without the isospin part are very small. In order to avoid a
long list of numbers we give in tab. \ref{d3}, for each nucleus
under investigation, the minimum, the maximum and the average
difference, in absolute value, between the calculated splittings. It
appears clear the relatively small importance of this term of the
interaction.
This fact can be understood considering that the major contribution
to the spin--orbit interaction is coming from the direct
${\cal U}^p_C$ term.
For the spin-orbit isospin term of
the interaction, the case $p=8$, the ${\cal U}^p_C$ term contains a
function which is given by the difference between the proton and
neutron density distributions, eq. (\ref{uuu}). 
In all the nuclei we have considered this difference is
small and particularly small in
those nuclei having the same number of protons and neutron ($^{16}$O
and $^{40}$Ca).  In effect the maximum
differences are larger in $^{48}$Ca and $^{208}$Pb than in $^{16}$O
and $^{40}$Ca.

\section{Summary and Conclusions}
In this article we have presented a formalism to treat finite range
spin--orbit interactions in HF calculations.  
The finite range of the interaction generates additional terms with
respect to the usual shell model expression. One of these is the
contribution to the exchange Fock--Dirac term in the HF
equation (\ref{hf6}).  Also in the direct (Hartree) term of this
equation there is a new part: the ${\cal U}_{LS}$ piece
of eq. (\ref{dls}). The major goal of our work was the 
investigation of the effects produced by these new components.
This has been done by adding different type of spin--orbit terms to a fixed
interaction active only in the four central channels. We have
used a spin--orbit interaction of a gaussian form whose parameters have been
fixed to reproduce the energy splitting of the proton $1p$ levels in
$^{16}$O.

We have shown that the largest part of the spin--orbit effects in
HF calculations is produced by the traditional shell model
term, ${\cal U}_{C}$ in eq. (\ref{dls}).  The
contribution of the other term, ${\cal U}_{LS}$, is very small and it
can be simulated by a redefinition of the parameters of the
force.
The role of the exchange term is more complicated: its inclusion in
the calculations modifies by a maximum of 15\% the values of the
spin--orbit splittings.  The complication arises because these
modifications do not have the same sign for all the nuclei studied.
In calculations done with only the direct terms, it is not possible to
simulate the exchange effects by simply readjusting the parameters of
the interaction.

Forcing our formalism to handle zero--range spin--orbit interactions
we have studied, by comparison, the importance of the finite range. We
found that calculations done with zero--range interaction produce
energy splittings which, in heavy nuclei, are much larger than the
empirical ones.
Traditional HF calculations use spin--orbit zero--range
terms of Skyrme type \cite{bel56}. These expressions produce
contributions to the hamiltonian which are related to the derivative
of the density distribution, while our expressions depend
directly from the density distribution. The dependence from the
derivative of the density distribution simulate effects produced
by the finite range of the force.

We have done calculations with spin-orbit terms taken from
microscopic interactions \cite{lag81,wir84,wir95} and we have obtained
splittings close to those produced by our effective spin--orbit
interactions. This would  indicate that the medium does not affect the
spin-orbit term of the realistic interaction, in agreement with the
findings of G-matrix calculations \cite{nak84}. We would like to point
out, however, that the observables we have investigated are more
sensitive to the global properties of the spin-orbit potential
than to its details. Modifications of the local properties of the
interaction would not produce effects on our results.

We have also investigated the effects of the spin--orbit isospin term
of the interaction, and we found them to be very small. These terms
are related to the differences between protons and neutrons density
and spin--orbit density distributions. In our calculations these
quantities are rather small even for a nucleus with a large neutron
excess like $^{208}$Pb.  There are however indications for observables
which seems to be sensitive to this part of the potential
\cite{lal98}.

The comparison of the results of our calculations with the empirical
values of the splittings on the various nuclei investigated is not
satisfactory. The empirical splittings are smaller than those we have
obtained, except for $^{12}$C. This is a well known problem of the
HF theory, and it could be solved only by using theories going beyond
the mean field description of the nucleus.

%----------------------------------------------------------------------
\section{Appendix}

Since we have developed the HF equations in 
spherical coordinates it is necessary to express the operator 
${\bf L}\cdot{\bf S}$ in terms of these coordinates. 
For this purpose we define an operator $O(ijk)$ as:
\begin{equation}
O(ijk)\equiv(-1)^{i+j}\,{\bf r}_i\times
{\bf p}_j\cdot{\bf s}_k=\sqrt{\frac{2\pi}{3}}(-1)^{i+j}r_i
\sum_{\mu}(-1)^{1-\mu}Y_{1\mu}(\widehat{r}_i)
\left[\nabla_j\otimes\sigma(k)\right]^{1}_{-\mu},
\end{equation}
\noindent
where we have set $\hbar=1$ and the indexes $i,j,k$ can assume only
two values, $1$ and $2$ for example.  The previous formula has been
obtained by expressing $r$ in terms of spherical harmonics and by
making explicit use of the tensor product properties. Using the above
operator we can write the spin--orbit channels of the force as:
\begin{eqnarray}
\label{potSO2}
V_p(r_{12})O_p(1,2)&\equiv & V_p(r_{12})\,\,\, \, 
{\cal I}^p\sum_{i,j,k=1,2} O(ijk) 
\nonumber
\\
&=& 4\pi \sqrt{\frac{2\pi}{3}}\,\,\,\, {\cal I}^p \, \sum_L
\frac{1}{{\widehat{L}}^2} {\cal V}^p_L(r_1,r_2) 
\sum_{jk} (-1)^j \sum_{\mu}(-1)^{1-\mu} 
\left[ \nabla_j \otimes \sigma(k) \right]^{1}_{-\mu} 
\nonumber
\\
&\mbox{}&\sum_{iM}(-1)^{M+i}\,r_i\,Y_{L-M}(\widehat{r}_1)\, 
Y_{LM}(\widehat{r}_2)\, Y_{1 \mu}
(\widehat{r}_i),\,\,\,\,\,\,\,\, \mbox{$p=7,8$} \, . 
\end{eqnarray}
\noindent
The operator ${\cal I}^p$ has been defined as:
\begin{equation}
{\cal I}^p\, =\, \left\{ \begin{array}{ll}
                 1, & \mbox{$p=7$} \\ \\
                 \btau(1)\cdot\btau(2), & \mbox{$p=8$} \, .
\end{array}
\right.
\end{equation}
In eq. (\ref{potSO2}) we make the sum on $M$ e $\mu$, and we obtain a
more synthetic expression:
\begin{eqnarray}
V_p(r_{12})O_p(1,2)&=&-2\sqrt{2\pi}\,{\cal I}^p
\,\sum_{LL'}\frac{\widehat{L'}}{\widehat{L}}\threej {L} {L'} {1} {0} {0} {0}
{\cal V}^p_L(r_1,r_2)
\nonumber 
\\
&\mbox{}& \; \sum_{ijk} O^{LL'}_{00}(ijk), \,\,\,\,\,\,\,\, p=7,8 \, ,
\end{eqnarray}
\noindent
where we have defined new set of operators as:
\begin{equation}
O^{LL'}_{00}(ijk)=(-1)^{i+j} r_i 
\left[[Y_{L'}(\widehat{r}_i)\otimes Y_L(\widehat{r}_{i\!\!/})]^1\otimes 
\left[\nabla_j\otimes\sigma(k)\right]^1\right]^0_0
\end{equation}
\noindent
and where we have defined:
\begin{equation}
{\bf r}_{i\!\!/}\,=\, \left\{ \begin{array}{ll}
                                {\bf r}_1, & \mbox{for \,\, $i$=2}\\
                                {\bf r}_2, & \mbox{for \,\, $i$=1} \, .
\end{array}
\right.
\end{equation}
It is convenient to express these operators such as the coordinates of
each particle are separated:
\begin{equation}
O^{LL'}_{00}(ijk)=\sqrt{3}\sum_{KM}(-1)^{L+M} 
\sixj {L} {L'} {1} {1} {1} {K} \tilde{O}^M_{LL'K}(ijk),
\end{equation}

\noindent
where we have used the Racah 6-$j$ symbol and we have 
defined the operators:
\begin{equation}
\tilde{O}^M_{LL'K}(ijk)=
e(ijk)\,\bar{A}^{\left(ijk\right)}_{JM}(1)\, 
\bar{B}^{\left(ijk\right)}_{J-M}(2).
\end{equation}
The expressions of the three terms of the above equation, for each
value of $(ijk)$ are given in tab. \ref{oper} as functions of the
following operators:
\begin{eqnarray}
\label{mathC}
{\cal C}^L_{\lambda\mu}(i)&=&\left[Y_{L}(\widehat{r}_i)
\otimes\nabla_i\right]^{\lambda}_{\mu} 
\nonumber
\\
{\cal S}^{LK}_{\lambda\mu}(i)&=&\left[\left[Y_{L}(\widehat{r}_i)
\otimes\nabla_i\right]^K \otimes\sigma(i)\right]^{\lambda}_{\mu} 
\equiv\left[{\cal C}^L_K(i)\otimes\sigma(i)\right]^{\lambda}_{\mu}\\
{\cal M}^L_{\lambda\mu}(i)&=&\left[Y_{L}(\widehat{r}_i) 
\otimes\sigma(i)\right]^{\lambda}_{\mu}.
\nonumber
\end{eqnarray}
At the end we express the spin--orbit terms of the interaction as: 
\begin{eqnarray*}
V_p(r_{12})O_p(1,2)&=& 4\pi\,{\cal I}^p \sum_{LL'K}
f(L,L',K)\,{\cal V}^p_L(r_1,r_2)\nonumber \\
&\mbox{}& \;\sum_{ijk}
\sum_M(-1)^M\,\tilde{O}^M_{LL'K}(ijk),\,\,\,\,\,\,\,\, p=7,8 \, , 
\end{eqnarray*}
\noindent
where  $f$ is given by:
\begin{equation}
f(L,L',K)=(-)^{L+K}\xi(L+L'+1)\frac{\widehat{L'}}{\widehat{L}}
\threej {1}{K}{L'}{1}{-1}{0}
\threej {1}{K}{L}{1}{-1}{0}.
\end{equation}

In the calculation of the HF equations for the
spin--orbit channels we have used the results corresponding to
the matrix elements  $\tilde{O}^M_{LL'K}(ijk)$ which in the tab.
\ref{oper} are shown to be function of
${\cal C}^K_{LM}$, ${\cal M}^K_{LM}$, ${\cal S}^{L'K}_{LM}$
defined in (\ref{mathC}) and of the spherical harmonics
$Y_{LM}(\widehat{r})$.

Using the function $\xi(l)$ =1 if $l$ is even and =0 if $l$ is odd,  
we express the reduced matrix elements for the spherical
harmonics as:

\begin{equation}
\langle l \half j || Y_L || l' \half j' \rangle\, =\, 
\frac{1}{\sqrt{4\pi}}(-1)^{j'+L+\frac{3}{2}}\,\widehat{j}\,\widehat{j'}\,
\widehat{L}\,\xi(l+l'+L)\,\threej{j}{j'}{L}{\half}{-\half}{0}.
\end{equation}

\noindent
For the other three operators we obtain the following operators: 
\begin{eqnarray}
\langle l \half j|| {\cal S}^{L'K}_L || l' \half j' \rangle &=& 
\sqrt{\frac{3}{2\pi}} (-1)^{l'}\,\widehat{j}\, 
\widehat{j'}\,\widehat{l}\,\widehat{l'}\,\widehat{L}\,\widehat{L'}\,\widehat{K}\,\ninej{l}
{\half}{j}{l'}{\half}{j'}{K}{1}{L}
\nonumber
\\
&\mbox{}&
\left\{\sqrt{2}\,\xi(l+l'+L'+1)\,\threej{l'}{K}{l}{1}{-1}{0}\threej{K}
{1}{L'}{1}{-1}{0}\,\frac{\sqrt{l'(l'+1)}}{r}
\nonumber
\right.
\\
&\mbox{}&\left.\,\,\,+ 
\threej{l'}{K}{l}{0}{0}{0}\threej{K}{1}{L'}{0}{0}{0}\,\frac{{\rm 
d}}{{\rm d}r}\right\},
\nonumber
\\\nonumber \\ \nonumber \\
\langle l \half j||\,\, {\cal C}^K_L \, || l' \half j' \rangle &=& 
\frac{1}{\sqrt{4\pi}} (-1)^{l+l'+\half+j'+L}\,\widehat{j}\, 
\widehat{j'}\,\widehat{l}\,\widehat{l'}\,\widehat{L}\,\widehat{K}\,\sixj{\
l}{j}{\half}{j'}{l'}{L}
\nonumber
\\
&\mbox{}& 
\left\{\sqrt{2}\,\xi(l+l'+K+1)\,\threej{l'}{L}{l}{1}{-1}{0}\threej{L}{
\
1}{K}{1}{-1}{0}\,\frac{\sqrt{l'(l'+1)}}{r}
\nonumber \right.
\\
&\mbox{}&\left.\,\,\,+\threej{l'}{L}{l}{0}{0}{0}\threej{L}{1}{K}{0}{0}
{0}\,\frac{{\rm d}}{{\rm d}r}\right\},
\nonumber
\\\nonumber \\  \nonumber \\                                          
         \langle l \half j|| {\cal M}^K_L\, || l' \half j' \rangle 
&=& \sqrt{\frac{3}{2\pi}} (-1)^l\,\widehat{j}\, 
\widehat{j'}\,\widehat{l}\,\widehat{l'}\,\widehat{L}\,\widehat{K}\,\ninej{\
l}{\half}{j}{l'}{\half}{j'}{K}{1}{L}\threej{l}{K}{l'}{0}{0}{0}.
\end{eqnarray}
%

%----------------------------------------------------------------------

%------------------------------------------------------------------
\newpage

\begin{table}
\caption{Parameters of the central force $B1a$ compared with the
original parameterization of the Brink and Boeker $B1$ force
\protect\cite{bri67}. The values of the $A$ are expressed in MeV,
while those of $b$ in fm$^{-2}$.}
\label{tb1b1a}
\begin{center}
\begin{tabular}{crr}
\hline
           & $B1$     & $B1a$     \\ 
\hline
 $A_{11}$    & -55.12 & -61.51  \\
 $A_{12}$    & 647.06 & 647.06  \\
 $A_{p1}$    &  17.10 &  15.82  \\ 
 $A_{p2}$    &  51.51 &  51.51  \\
 $b_1$       &   0.51 &   0.51  \\
 $b_2$       &   2.04 &   2.04  \\
\hline
\end{tabular}
\end{center}
\end{table}
%

%
%\begin{table}
%\caption{Parameters of the spin-orbit terms of the force.}
%\label{tforce}
%\begin{center}
%\begin{tabular}{crc}
%\hline
%           & $A_7 [MeV]$ & $b_7 [fm^{-2}]$ \\ 
%\hline
% $z$         & -7.50 &        \\
% $c$        & -108.75 & 1.2  \\
% $d$         & -107.50 & 1.2  \\
% $so$        & -97.86  & 1.2  \\ 
% d$\tau$   & -107.5 & 1.2  \\
% so$\tau$  & -97.9  & 1.2   \\  
%\hline
%\end{tabular}
%\end{center}
%X\end{table}
%
%
\begin{table}
\caption{Binding energies, in MeV, for the nuclei investigated.
The force $B1a$ does not have spin--orbit terms, the 
$z$, $c$, $d$ and $so$ calculations are discussed in the text.
 The interactions labeled $U$, $A14$ and
$A18$ have been obtained by adding to $B1a$ the
spin-orbit terms of the Urbana V14 potential \protect\cite{lag81}, of
the Argonne V14 \protect\cite{wir84} potential and the Argonne V18
\protect\cite{wir95} potential. The experimental energies are taken
from ref.~\protect\cite{aud93}. }
\label{be}
\begin{center}
\begin{tabular}{cccccc}
\hline
       & $^{12}$C & $^{16}$O & $^{40}$Ca & $^{48}$Ca & $^{208}$Pb  \\
\hline
%$B1$      & -48.68  & -113.55  & -340.75   & -362.82   &  -2059.55 \\
$B1a$     & -48.89  & -116.09  & -341.54   & -380.48   &  -1854.00 \\
$z$       & -54.47  & -116.08  & -341.43   & -402.67   &  -2053.15 \\
$c$       & -55.26  & -116.09  & -341.52   & -397.74   &  -1913.47 \\
$d$       & -63.13  & -116.42  & -342.30   & -413.36   &  -1985.81 \\
$so$      & -67.37  & -116.42  & -342.24   & -420.60   &  -2007.25 \\
$U$       & -62.77  & -116.46  & -342.27   & -411.75   &  -1969.91 \\
$A14$    & -55.18  & -116.19  & -341.72   & -395.20   &  -1907.95 \\
$A18$    & -61.20  & -116.37  & -342.10   & -408.55   &  -1960.31 \\
exp       & -92.16  & -127.62  & -342.05   & -416.00   &  -1636.45 \\
\hline
\end{tabular}
\end{center}
\end{table}
\begin{table}
\caption{Spin-orbit splittings in MeV 
for the nuclei investigated.  The empirical values
have been calculated using the compilation of
ref.~\protect\cite{cam72}.  $\pi$ stands for protons, $\nu$ for
neutrons.}
\label{tz1}
\begin{center}
\begin{tabular}{cccccccccc}
\hline
         &  $z$  &  $c$  & $d$   & $so$    
                 &  U    & A14   & A18   & SIII & exp   \\
\hline
$^{12}$C &       &       &       &       
                 &       &       &       &      &   \\
 1p$\pi$ &  3.37 &  3.90 &  4.97 &  5.76 
                 &  4.88 &  1.58 &  4.16 &      &  14.00  \\  
 1p$\nu$ &  3.05 &  3.57 &  4.65 &  5.46 
                 &  4.61 &  1.24 &  3.83 &      &  13.76  \\  
\hline
$^{16}$O &       &       &       &       
                 &       &       &       &       &         \\
 1p$\pi$ &  6.30 &  6.31 &  6.31 &  6.32 
                 &  5.78 &  2.86 &  5.33 &  5.90 &  6.33  \\
 1p$\nu$ &  6.38 &  6.36 &  6.36 &  6.37 
                 &  5.89 &  2.91 &  5.41 &  6.03 &  6.16  \\
\hline
$^{40}$Ca &      &       &       &       
                 &       &       &       &       &        \\
 1p$\pi$ &  9.38 &  5.56 &  5.64 &  5.35 
                 &  4.91 &  2.40 &  4.59 &  3.31 &        \\
 1d$\pi$ & 12.14 &  8.36 &  8.45 &  9.16 
                 &  7.52 &  3.70 &  6.99 &  6.18 &  7.17  \\
 1p$\nu$ &  9.47 &  5.58 &  5.66 &  5.38 
                 &  5.03 &  2.47 &  4.67 &  3.38 &  3.00  \\
 1d$\nu$ & 12.43 &  8.46 &  8.56 &  9.30 
                 &  7.74 &  3.82 &  7.15 &  6.33 &  6.30  \\
\hline
$^{48}$Ca &      &       &       &       
                 &       &       &       &       &   \\
 1p$\pi$ & 10.05 &  4.62 &  5.11 &  4.94 
                 &  4.47 &  2.10 &  4.13 &  2.72 &        \\
 1d$\pi$ & 13.97 &  7.50 &  8.27 &  9.03 
                 &  7.52 &  3.70 &  6.99 &  6.18 &  4.30  \\
 1p$\nu$ &  9.25 &  3.89 &  4.35 &  3.76 
                 &  3.41 &  1.23 &  3.20 &  2.36 &        \\
 1d$\nu$ & 12.68 &  6.35 &  7.09 &  7.49 
                 &  5.98 &  2.21 &  5.47 &  5.50 &  3.60  \\
 1f$\nu$ & 11.69 &  7.82 &  8.53 & 10.17 
                 &  8.07 &  3.28 &  7.16 &      &  8.74  \\
\hline
\end{tabular}
\end{center}
\end{table}
\begin{table}
\caption{The same as tab. \protect\ref{tz1} for $^{208}$Pb.}
\label{tz2}
\begin{center}
\begin{tabular}{cccccccccc}
\hline
         &  $z$  &  $c$  & $d$   & $so$    
                 &  U    & A14   & A18   & SIII & exp   \\
\hline
$^{208}$Pb &     &       &       &        
                 &       &       &       &     \\
 1p$\pi$ & 14.61 &  1.98 &  2.05 &  1.60 
                 &  1.61 &  0.67 &  1.54 &  0.52  &        \\
 1d$\pi$ & 22.32 &  3.36 &  3.63 &  3.56 
                 &  2.79 &  1.08 &  2.66 &  1.25  &        \\
 1f$\pi$ & 28.47 &  5.01 &  5.54 &  5.86 
                 &  4.33 &  1.67 &  4.10 &  2.34  &        \\
 1g$\pi$ & 32.76 &  6.97 &  7.77 &  8.56 
                 &  6.26 &  2.53 &  5.88 &  3.76  &        \\
 1h$\pi$ & 34.31 &  9.15 & 10.17 & 11.47 
                 &       &       &       &        &  5.57  \\
 2p$\pi$ & 12.55 &  2.34 &  2.45 &  1.96 
                 &  2.12 &  0.91 &  1.96 &  0.99 &        \\
 2d$\pi$ & 18.78 &  4.00 &  4.18 &  4.29 
                 &  3.60 &  1.63 &  3.36 &  1.76  &  1.32  \\
 2f$\pi$ & 21.03 &  5.23 &  5.46 &  5.98 
                 &  4.77 &  2.00 &  4.41 &  2.35  &  1.92  \\
 1p$\nu$ & 14.63 &  2.05 &  2.11 &  1.46 
                 &  1.80 &  0.83 &  1.68 &  0.53 &        \\
 1d$\nu$ & 22.44 &  3.48 &  3.74 &  3.55 
                 &  3.17 &  1.38 &  2.94 &  1.26 &        \\
 1f$\nu$ & 28.63 &  5.10 &  5.64 &  5.83 
                 &  4.77 &  2.00 &  4.41 &  2.35 &        \\
 1g$\nu$ & 32.85 &  6.96 &  7.77 &  8.38 
                 &  6.67 &  2.79 &  6.13 &  3.81 &        \\
 1h$\nu$ & 34.06 &  8.89 &  9.92 & 11.08 
                 &  8.79 &  3.75 &  8.00 &  5.56 &        \\
 2p$\nu$ & 12.50 &  2.31 &  2.43 &  1.58 
                 &  2.08 &  0.89 &  1.93 &  1.08 &        \\
 2d$\nu$ & 18.74 &  3.92 &  4.10 &  3.96 
                 &  3.73 &  1.68 &  3.40 &  1.98  &        \\
 2f$\nu$ & 20.94 &  5.15 &  5.36 &  5.97 
                 &  5.17 &  2.41 &  4.64 &  2.78  &  1.77  \\
 3p$\nu$ &  8.05 &  2.01 &  2.01 &  1.52 
                 &  1.95 &  0.92 &  1.76 &  1.03 &  0.94  \\
\hline
\end{tabular}
\end{center}
\end{table}
\begin{table}
\caption{Differences in MeV between the splittings presented in tabs.
\protect\ref{tz1} and \protect\ref{tz2} and those obtained leaving out
the isospin part of the spin--orbit interaction for Urbana, Argonne V14
and Argonne V18 potentials. We give here the minimum, maximum and
average difference, in absolute value.}
\label{d3}
\begin{center}
\begin{tabular}{cccc}
\hline
           & Minimum &  Maximum  & Average   \\
\hline
$^{16}$O   & 0.010    & 0.030      & 0.018       \\
$^{40}$Ca  & 0.000    & 0.070      & 0.032      \\
$^{48}$Ca  & 0.020    & 0.290      & 0.088      \\
$^{208}$Pb & 0.000    & 0.450      & 0.128       \\
\hline
\end{tabular}
\end{center}
\end{table}
\begin{table}
\caption{\label{oper} 
Expressions of the $\tilde{O}^M_{LL'K}(ijk)$ operators.}
\begin{center}
\begin{tabular}{c|ccc}
\hline
\hline
\rule[-2mm]{0mm}{8mm}$i\,j\,k$ & $e(ijk)$ &
$\bar{A}^{\left(ijk\right)}_{JM}$ 
& $\bar{B}^{\left(ijk\right)}_{J-M}$  \\ \hline
\hline
\rule[-3mm]{0mm}{8mm}1\,1\,1 & $-r_1\frac{\widehat{K}}{\widehat{L}}$ & ${\cal
S}^{L'K}_{LM}$ & $Y_{L-M}$ \\ \hline
\rule[-3mm]{0mm}{8mm}1\,1\,2 & $r_1$                & ${\cal C}^{L'}_{KM}$  &
${\cal M}^L_{K-M}$ \\  \hline
\rule[-3mm]{0mm}{8mm}1\,2\,1 & $r_1$                & ${\cal M}^{L'}_{KM}$ &
${\cal C}^L_{K-M}$  \\  \hline                      
\rule[-3mm]{0mm}{8mm}1\,2\,2 & $-r_1\frac{\widehat{K}}{\widehat{L'}}$ &
$Y_{L'M}$ & 
${\cal S}^{LK}_{L'-M}$ \\ \hline
\rule[-3mm]{0mm}{8mm}2\,1\,1 & $-r_2\frac{\widehat{K}}{\widehat{L'}}$ & ${\cal
  S}^{LK}_{L'M}$ & $Y_{L'-M}$ \\ \hline
\rule[-3mm]{0mm}{8mm}2\,1\,2 & $r_2$                & ${\cal  C}^L_{KM}$  
& ${\cal M}^{L'}_{K-M}$ \\ \hline
\rule[-3mm]{0mm}{8mm}2\,2\,1 & $r_2$                & ${\cal  M}^L_{KM}$ 
& ${\cal C}^{L'}_{K-M}$  \\ \hline
\rule[-3mm]{0mm}{8mm}2\,2\,2 & $-r_2\frac{\widehat{K}}{\widehat{L}}$ &$Y_{LM}$ 
& ${\cal S}^{L'K}_{L-M}$ \\ \hline
\hline
\end{tabular}
\end{center}
\end{table}

\newpage

\begin{figure}[ht]
\vspace{-2.5cm}
\begin{center}                                                                
\leavevmode
\epsfysize = 500pt
\hspace*{.45cm}
\makebox[0cm]{\epsfbox{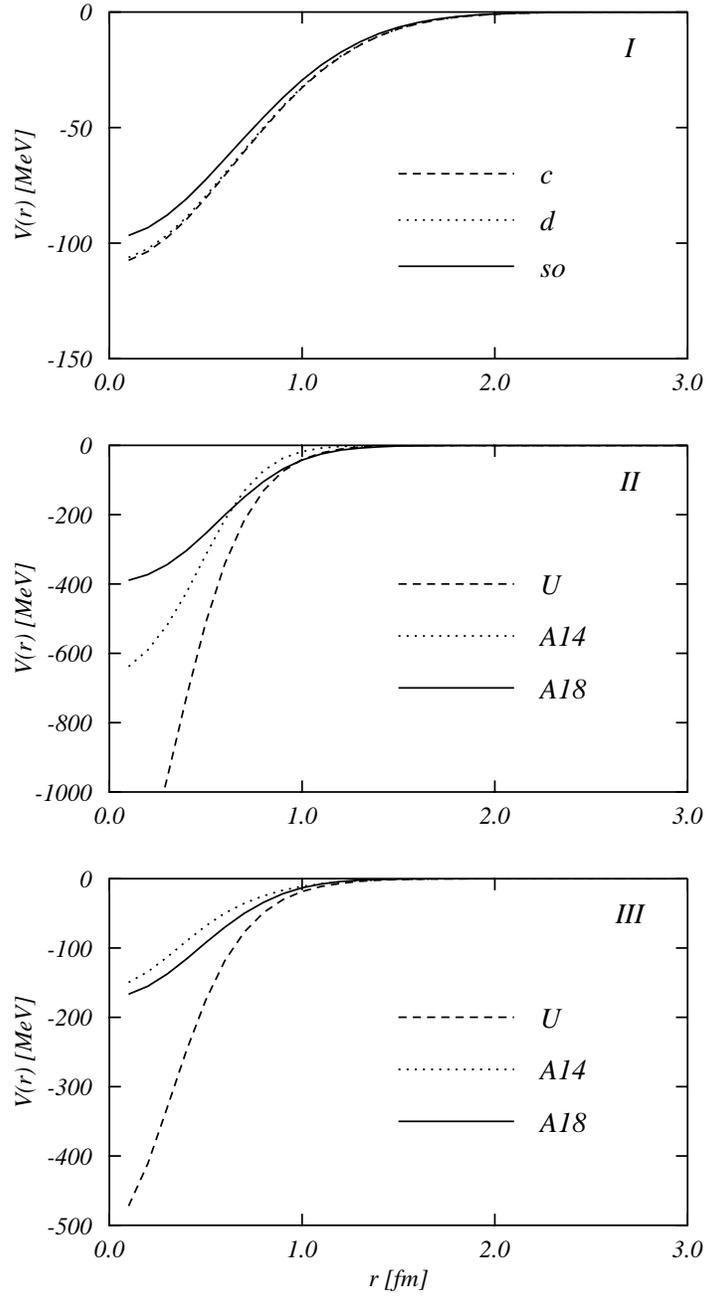}}
\end{center}
\vspace{1.cm}
\caption[]{ \small Radial dependence of the spin--orbit interaction used.
 In the panel I our gaussian interactions are shown while 
 in the other two panels the radial dependence of the spin--orbit
 terms (II) and  spin--orbit isospin terms (III)  of the Urbana
 \protect\cite{lag81}  Argonne V14  \protect\cite{wir84} 
 and Argonne V18  \protect\cite{wir95} interactions. }
\label{figu}
\end{figure}

\end{document}